\documentclass[12pt]{article}
\usepackage{amsmath}
\usepackage{amssymb}
\usepackage{graphicx}
\usepackage{float}
\usepackage{indentfirst}
\usepackage{mathrsfs}
\usepackage{braket}
\usepackage{dsfont}%
\usepackage{multirow}
\usepackage{diagbox}
\usepackage{color}
\usepackage{hyperref}					
\usepackage[capitalise,noabbrev]{cleveref}

\usepackage{setspace}

\usepackage[labelfont=bf,labelsep=period]{caption}

\usepackage[numbers,sort&compress]{natbib}

\newtheorem{theorem}{Theorem}

\newtheorem{lemma}{Lemma}

\title{The Decompositions of Werner and Isotropic States \\ [0.8cm] }

\author
{Ma-Cheng Yang,$^1$ Jun-Li Li,$^{2,3}$ and Cong-Feng Qiao$^{1,3}\footnote{Corresponding author, qiaocf@ucas.ac.cn}$ \\ [0.2cm]
\normalsize{$^1$School of Physical Sciences, University of Chinese Academy of Sciences,} \\
\normalsize{YuQuan Road 19A, Beijing 100049, China}\\ [2pt] \normalsize{$^2$Center of Materials Science and Optoelectronics Engineering \& CMSOT,} \\
\normalsize{University of Chinese Academy of Sciences, YuQuan Road 19A, Beijing 100049, China}\\ [2pt]
\normalsize{$^3$Key Laboratory of Vacuum Physics, University of Chinese Academy of Sciences} \\
\normalsize{YuQuan Road 19A, Beijing 100049, China} \\ [3mm]
}

\date{}

\begin{document}
\baselineskip24pt \maketitle
\begin{abstract} \doublespacing
The decompositions of separable Werner state, and also isotropic state, are well-known tough issues in quantum information theory, in this work we investigate them in the Bloch vector representation, exploring the symmetric informationally complete positive operator-valued measure (SIC-POVM) in the Hilbert space. We successfully get the decomposition for arbitrary $N\times N$ Werner state in terms of regular simplexes. Meanwhile, the decomposition of isotropic state is found to be related to the decomposition of Werner state via partial transposition. It is interesting to note that in the large $N$ limit, while the Werner states are either separable or non-steerably entangled, most of the isotropic states tend to be steerable.
\end{abstract}

\newpage

\section{Introduction}

As one of the essential physical resources in quantum information processing, entanglement has been studied extensively, while we are still in the early stage of fully understanding it \cite{Ent-rev}. The essence of the quantum entanglement is the inherent nonlocal correlation that is fundamentally different from the classical situation. The quantum nonlocality, which violates the local realistic theory, has experienced a large number of various experimental tests \cite{Bell-1, Bell-13}, and is as of yet well established.

Werner proposed an important class of states, the Werner state, for the study the separability of mixed state \cite{Werner-state}, and was found that the non-separability of it is not enough to suffice the Bell non-locality. The Werner state bears high symmetry, and hence its many properties can be conveniently investigated for the aim of quantum information processing, such as steering \cite{Steer-1}, quantum discord \cite{Quan-Corr}, distillability \cite{Bound-ent1,Bound-ent2}, etc.

The Werner state has been investigated with great effort, however, till now we still do not have the explicit decomposition form for arbitrary separable Werner state. Refs. \cite{Ent-for,Ano-con,Werner-Wigner,Quan-des} discover some solutions applicable merely to specific separable states. In Ref. \cite{Separability-1} the decomposition of the separable Werner state was investigated, and authors conjectured that the Werner state may be decomposed by regular simplex with circumradius $\sqrt{\frac{2(N-1)}{N}}$ in the Bloch-vector space of $N$-dimensional mixed state. We know the existence of the regular simplex in this case is equivalent to the existence of a symmetric, informationally complete, positive operator-valued measure (SIC-POVM), and the latter has been widely studied \cite{SIC-QM}.

In this paper, parameterized with one single parameter in the Bloch vector space, we decompose all the Werner and isotropic states, not necessarily separable, into two complementary regular simplexes. The decomposition of the separable Werner and isotropic states are found can be constructed on condition that the SIC-POVM exists, where the local density matrices are obtained from the simplex formed by SIC-POVM. We will show as well that in large dimension limit of $N$ the Werner states tend to be either separable or non-steerably entangled, while the isotropic states appear to be all steerable as the dimension grows.

\section{The decomposition of Werner and isotropic states}

\subsection{The Bloch representation of quantum states}

An arbitrary $N$-dimensional density matrix $\rho$ may be represented as follows:
\begin{align}
\rho = \frac{1}{N}\mathds{1} + \frac{1}{2}\sum_{\mu=1}^{N^2-1} r_{\mu} \lambda_{\mu} = \frac{1}{N}\mathds{1} + \frac{1}{2}\vec{r} \cdot \vec{\lambda}\; , \label{a-2}
\end{align}
where $\lambda_{\mu}$ are $N^2-1$ generators of SU($N$) group and $\vec{r}$ is a $(N^2-1)$-dimensional Bloch vector with component $r_{\mu} = \mathrm{Tr}[\rho \lambda_{\mu}]$. The density matrix $\rho$ needs to be positive semi-definite and trace one, which impose constraints on the Bloch vector $\vec{r}$ \cite{Bloch-N-systems, Bloch-positivity}. For example, the norm of the vector $|\vec{r}\, | \leq \sqrt{2(N-1)/N}$ since $\mathrm{Tr}[\rho^2]\leq 1$. Considering the positivity of density matrix, we have the following Lemma:
\begin{lemma}
For trace one Hermitian matrix $\displaystyle \rho =\frac{1}{N}\mathds{1} + \frac{1}{2}r \hat{r}\cdot \vec{\lambda}$, where $\hat{r}$ is a unit vector, the following statements are true:\\
(1) If $\displaystyle \rho' = \frac{1}{N}\mathds{1} + \frac{1}{2}s \hat{r}\cdot \vec{\lambda}$ has the Bloch vector opposite to $\rho$, i.e., $s <0$, the positive semidefiniteness of $\rho$ and $\rho'$ gives
\begin{align}
rs \geq \frac{-2}{N} \; .
\end{align}
(2) Given $\rho = \frac{1}{N}\mathds{1} + \frac{1}{2}r \hat{r}\cdot \vec{\lambda}$ a pure state at $r=\sqrt{2(N-1)/N}$, it will be positive semidefinite if and only if
\begin{align}
-\sqrt{\frac{2}{N(N-1)}} \leq r\leq \sqrt{\frac{2(N-1)}{N}} \;.
\end{align} \label{Lemma-opposite}
\end{lemma}
\noindent{\bf Proof:} 1. Based on Corollary 7.6.2. of Ref. \cite{Matrix-Book}, two positive semidefinite matrices satisfy
\begin{align}
\mathrm{Tr}[\rho\rho'] = \frac{1}{N} + \frac{1}{2} rs \geq 0  \; .
\end{align}
2. The necessary and sufficient condition is obvious in light of the Observation 4 of Ref. \cite{Separability-1} and as per the convexity nature of density matrix space. Q.E.D.

\subsection{Werner and isotropic states in different representations}

By definition, the Werner states satisfy $\rho_{\mathrm{W}} = (u\otimes u)\rho_{\mathrm{W}}(u^{\dag}\otimes u^{\dag})$ with $u$ a unitary matrix, and in $N\times N$ dimensional Hilbert space it may be constructed as \cite{H-Reduction-1}
\begin{align}
\rho_{\mathrm{W}} = \left( \frac{N-\phi}{N^3-N}\right)\mathds{1} \otimes \mathds{1} + \left(\frac{N\phi -1}{N^3-N} \right)  \mathds{V}  \; ,
\end{align}
where, $\phi$ is an arbitrary number $\in [-1,\ 1]$, $\mathds{V}$ is defined by $\mathds{V}\ket{\psi}\otimes \ket{\varphi} = \ket{\varphi} \otimes \ket{\psi}, \forall \ket{\psi},\ket{\varphi} \in \mathcal{H}$, the $N$-dimensional Hilbert space. It is known that $\rho_{\mathrm{W}}$ is separable if and only if $\phi>0$ \cite{Werner-state}. In the study of entanglement distillation, the Werner states are usually parameterized as \cite{Werner-distilla-form}
\begin{align}
\rho_{\mathrm{W}} = \frac{1}{N^2+\alpha N}\left(\mathds{1} \otimes \mathds{1} + \alpha \mathds{V}\right) \;  .
\end{align}
Here $\alpha = \frac{N\phi -1}{N-\phi}\in [-1,1]$. While considering of the steerability, Werner states are more convenient to be in such a representation as \cite{Steer-1}
\begin{align}
\rho_{\mathrm{W}} =  \left(\frac{N-1+ \beta}{N^3-N^2} \right) \mathds{1} \otimes \mathds{1} - \left(\frac{\beta}{N^2-N} \right)  \mathds{V} \;,
\end{align}
where $\beta = \frac{1-N\phi}{N+1}\in [\frac{1-N}{N+1},1]$.

By definition, the isotropic states are those states satisfying $\rho_{\mathrm{I}} = (u\otimes u^*) \rho_{\mathrm{I}} (u^{\dag} \otimes u^{*\dag})$, and generally parameterized in a symmetric form \cite{H-Reduction-1}
\begin{align}
\rho_{\mathrm{I}} = \frac{1-\eta}{N^2} \mathds{1}\otimes \mathds{1} + \eta P_{+} \; .
\end{align}
Here $\frac{-1}{N^2-1} \leq \eta\leq 1$ and $P_+ = |\psi_+\rangle\langle \psi_+|$ with $|\psi_+\rangle = \frac{1}{\sqrt{N}}\sum_{i}|ii\rangle$. The isotropic states turn out to be entangled if and only if $\frac{1}{N+1}<\eta$ \cite{H-Reduction-1}, while it will be steerable if and only if $\frac{H_N-1}{N-1} <\eta \leq 1$ with $H_N=\sum_{n=1}^N \frac{1}{n}$ the Harmonic number  \cite{Steer-1}.

\begin{figure}\centering
\scalebox{0.6}{\includegraphics{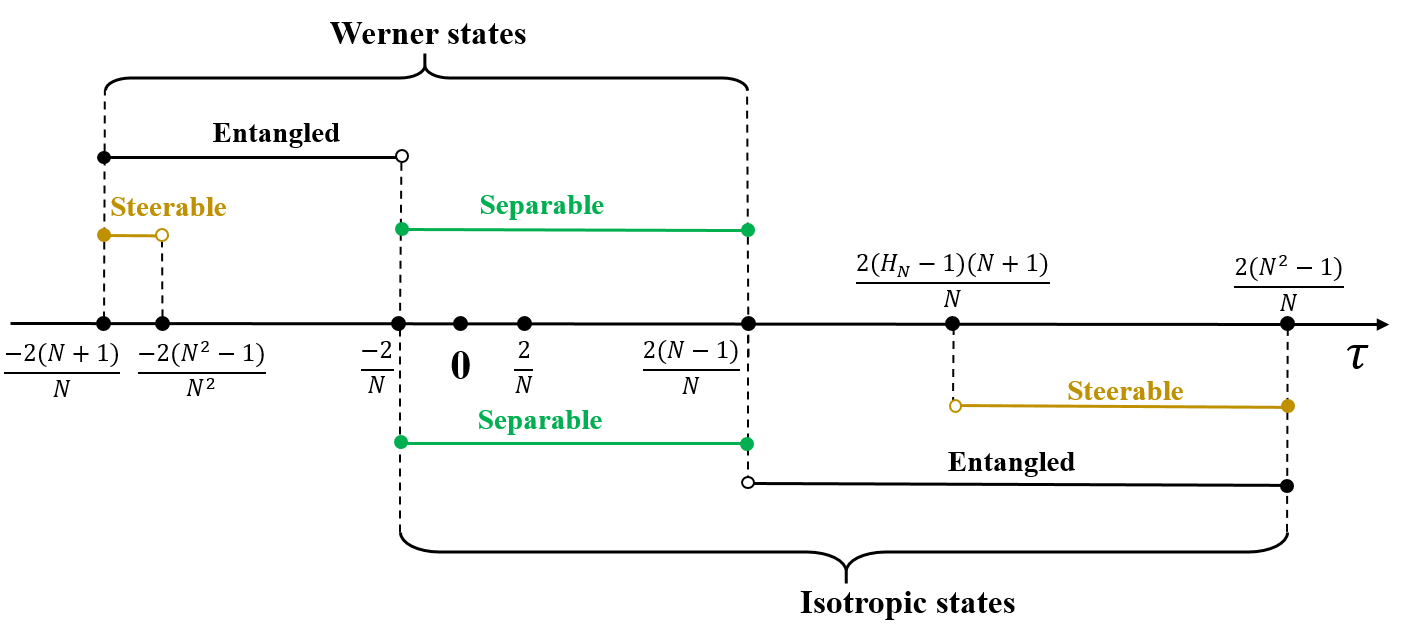}}
\caption{{\bf The nonlocalities in Werner and isotropic states.} Different types of non-localities emerge with different values of $\tau$ for Werner states $\rho_{\mathrm{W}}$ and isotropic states $\rho_{\mathrm{I}} $, where the solid and hollow circles mean the closed and open interval respectively, and $H_N$ denotes the $N$-th Harmonic number. } \label{Figure-WI-parameters}
\end{figure}

In Bloch representation, Werner states and isotropic states can be reformulated as \cite{Separability-Bloch}
\begin{align}
\rho_{W} & = \frac{1}{N^2} \mathds{1} \otimes \mathds{1} + \frac{1}{4}\sum_{\mu=1}^{N^2-1} \frac{2(N\phi-1)}{N(N^2-1)} \lambda_{\mu} \otimes \lambda_{\mu}\; ,
\end{align}
and
\begin{align}
\rho_{\mathrm{I}} = \frac{1}{N^2} \mathds{1}\otimes \mathds{1} + \frac{1}{4} \sum_{\mu=1}^{N^2-1} \frac{2\eta}{N} \lambda_{\mu} \otimes \lambda_{\mu}^{\mathrm{T}} \; ,
\end{align}
respectively, where superscript $\mathrm{T}$ signifies the transpose operation of a matrix. To investigate the nonlocal nature of these two classes of quantum states uniformly, we reparameterize the Werner states and isotropic states as
\begin{align}
\rho_{\mathrm{W}} & =  \frac{1}{N^2} \mathds{1} \otimes \mathds{1} + \frac{1}{4}\sum_{\mu=1}^{N^2-1} \left(\frac{\tau}{N^2-1}\right) \lambda_{\mu} \otimes \lambda_{\mu} \; , \label{Werner-tau}\\
\rho_{\mathrm{I}} & =  \frac{1}{N^2} \mathds{1} \otimes \mathds{1} + \frac{1}{4}\sum_{\nu=1}^{N^2-1} \left( \frac{\tau}{N^2-1} \right) \lambda_{\nu} \otimes \lambda_{\nu}^{\mathrm{T}} \; . \label{ISO-tau}
\end{align}
Here the parameter $\tau$ relates to $\phi$ and $\eta$ in Werner and isotropic states as
\begin{eqnarray}
\rho_{\mathrm{W}}:\; \tau = \frac{2(N\phi -1)}{N} \; ; \;\ \
\rho_{\mathrm{I}}:\; \tau = \frac{2\eta(N^2-1)}{N} \; .
\end{eqnarray}
According to the existing nonlocal criteria for $\rho_{\mathrm{W}}$ and $\rho_{\mathrm{I}}$, one may readily find the corresponding conditions for parameter $\tau$, see Figure \ref{Figure-WI-parameters}.

\subsection{Decomposition via the regular simplex}

A set of $(N^2-1)$-dimensional unit vectors $\mathcal{A} = \{\vec{a}_i|i=1,\cdots, N^2; |\vec{a}_i|  = 1 \}$ form a regular simplex, $N^2$-simplex, if
\begin{align}
\vec{a}_i\cdot \vec{a}_j = -\frac{1}{N^2-1} \; , \forall i\neq j \; .
\end{align}
For Werner and isotropic states paramterized by $\tau$ in \cref{Werner-tau,ISO-tau}, the following Lemma exists:
\begin{lemma}
For arbitrary $N^2$-simplex $\mathcal{A}$ in dimension $N^2-1$, the Werner state $\rho_W$ can always be decomposed as
\begin{align}
\rho_{\mathrm{W}} & = \sum_{i=1}^{N^2} \frac{1}{N^2} R_i(r) \otimes S_i(s) \; , \label{rhoW-Expand}
\end{align}
where $rs =\tau $, and
\begin{align}
R_i(r) = \frac{1}{N}\mathds{1} + \frac{r}{2} \vec{a}_i\cdot \vec{\lambda}\; , \; S_i(s) =\frac{1}{N}\mathds{1} + \frac{s}{2}\vec{a}_i\cdot \vec{\lambda} \; ,
\end{align}
are two trace one Hermitian matrices whose Bloch vectors form two $N^2$-simplexes of size $r$ and $s$, respectively. For isotropic states,
\begin{align}
\displaystyle  \rho_{\mathrm{I}} = \sum_{i=1}^{N^2} \frac{1}{N^2} R_i(r) \otimes S^{\mathrm{T}}_i(s)\ .
\end{align}
Here $R_i$ and $S_i$ may not be positive semidefinite. \label{Lemma-decom-RS}
\end{lemma}
\noindent{\bf Proof:} For $\vec{a}_i\in \mathcal{A}$, it is easy to verify that the following matrix is real orthogonal
\begin{align}
O = \sqrt{\frac{N^2-1}{N^2}} \begin{pmatrix}
\vec{a}_1 &  \vec{a}_2 & \cdots & \vec{a}_{N^2} \\
\frac{1}{\sqrt{N^2-1}} & \frac{1}{\sqrt{N^2-1}} & \cdots & \frac{1}{\sqrt{N^2-1}}
\end{pmatrix} \; ,
\end{align}
where $\vec{a}_i$ are $(N^2-1)$-dimensional column vectors. The relation $OO^{\mathrm{T}}= \mathds{1}$ tells
\begin{align}
\frac{N^2-1}{N^2}
\begin{pmatrix}
\vec{a}_1 &  \vec{a}_2 & \cdots & \vec{a}_{N^2} \\
\frac{1}{\sqrt{N^2-1}} & \frac{1}{\sqrt{N^2-1}} & \cdots & \frac{1}{\sqrt{N^2-1}}
\end{pmatrix}
\begin{pmatrix}
\vec{a}_1^{\,\mathrm{T}} & \frac{1}{\sqrt{N^2-1}} \\
\vec{a}_2^{\,\mathrm{T}} & \frac{1}{\sqrt{N^2-1}} \\ \\
\vdots & \vdots \\
\vec{a}_{N^2}^{\,\mathrm{T}} & \frac{1}{\sqrt{N^2-1}}
\end{pmatrix} = \mathds{1} \; . \label{Orth-MatA}
\end{align}
Here $\displaystyle \sum_{i=1}^{N^2} \vec{a}_i \vec{a}_i^{\mathrm{T}} = \frac{N^2}{N^2-1} \mathds{1}$ is an $(N^2-1)\times (N^2-1)$ identity matrix, and $\displaystyle \sum_{i=1}^{N^2}\vec{a}_i =0$. Hence \cref{rhoW-Expand} becomes
\begin{align}
\rho_{\mathrm{W}} & = \frac{1}{N^2} \mathds{1} \otimes \mathds{1} + \frac{rs}{4N^2} \sum_{i=1}^{N^2} \vec{a}_i\cdot\vec{\lambda} \otimes \vec{a}_i \cdot \vec{\lambda} \nonumber \\
& = \frac{1}{N^2} \mathds{1} \otimes \mathds{1} + \frac{rs}{4N^2} \sum_{\nu,\mu=1}^{N^2-1} T_{\mu\nu} \lambda_{\mu} \otimes \lambda_{\nu} \; ,
\end{align}
where $T_{\mu\nu} = \displaystyle \sum_{i=1}^{N^2} \vec{a}_i \vec{a}_i^{\mathrm{T}} = \frac{N^2}{N^2-1} \delta_{\mu\nu}$ is a diagonal matrix, and $\tau = rs$. The $\rho_{\mathrm{I}}$ would be similarly decomposed by taking the partial transposition.
Q.E.D.

\subsection{The separable decomposition by SIC-POVM}

\begin{figure}\centering
\scalebox{0.55}{\includegraphics{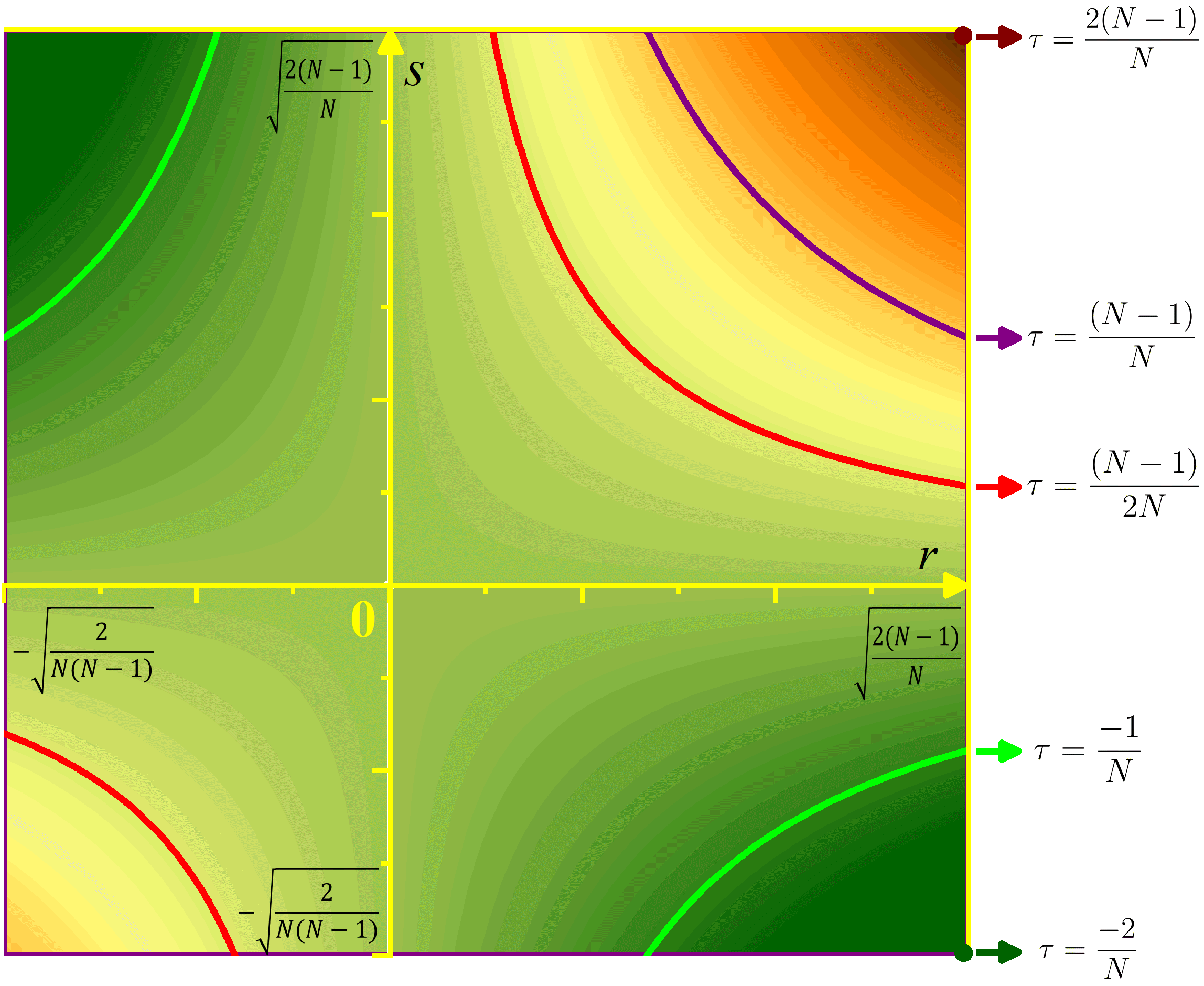}}
\caption{{\bf The separable decomposition of Werner states.} Infinite number of decompositions for each $\tau \in [-\frac{2}{N},\frac{2(N-1)}{N}]$ exist, except the endpoints of $-\frac{2}{N}$ and $\frac{2(N-1)}{N}$. The decompositions are achieved by the values of $r$ and $s$ fitting the contour $rs=\tau$.} \label{Figure-contour}
\end{figure}

A SIC-POVM in $N$-dimensional Hilbert space is represented by a set of $N^2$ vectors $|\psi_i\rangle$ satisfying \cite{SIC-QM}
\begin{align}
|\langle \psi_i|\psi_j\rangle |^2 = \frac{N\delta_{ij}+1}{N+1}\; , \; i,j\in \{1,\cdots, N^2\} \; . \label{Sic-states}
\end{align}
Let $\vec{r}_{i}$ and $\vec{r}_{j}$ be Bloch vectors of $\rho_i=|\psi_i\rangle \langle \psi_i|$ and $\rho_j= |\psi_j\rangle \langle \psi_j|$ respectively, then from \cref{Sic-states} we have
\begin{align}
\frac{\vec{r_{i}} \cdot \vec{r_{j}}}{|\vec{r}_i| |\vec{r}_j|} = -\frac{1}{N^2-1}\; , \; \forall i\neq j \; .
\end{align}
That is, the $N^2$ vectors $\vec{r}_{i}$ from SIC-POVM configures an $N^2$-simplex in the $N^2-1$ dimensional Bloch vector space. Therefore, we have the following theorem
\begin{theorem}
The separable Werner states, i.e. $\frac{-2}{N} \leq \tau \leq \frac{2(N-1)}{N}$, can always be decomposed into the following form
\begin{align}
\rho_{\mathrm{W}} & = \sum_{i=1}^{N^2} \frac{1}{N^2} R_i(r) \otimes S_i(s)   \;  \label{Th1-equ-W}
\end{align}
with
\begin{align}
R_{i}(r) = \frac{1}{N}\mathds{1} + \frac{r}{2 |\vec{r}_i|} \vec{r}_i \cdot \vec{\lambda} \; , \; S_{i}(s) = \frac{1}{N}\mathds{1} + \frac{s}{2|\vec{r}_i|} \vec{r}_i \cdot \vec{\lambda} \; .\nonumber
\end{align}
Here, $rs= \tau  \in [\frac{-2}{N},\frac{2(N-1)}{N}]$ and $r,s\in \left[-\sqrt{\frac{2}{N(N-1)}} ,\sqrt{\frac{2(N-1)}{N}}\right]$. \label{Theorem-sep}
\end{theorem}
\noindent{\bf Proof:}
In \cref{Lemma-decom-RS}, we have proved that the Werner and isotropic states can be decomposed by $R_{i}(r)$ and $S_{i}(s)$, whose Bloch vectors correspond to a regular simplex. Thus, we only need to prove that there exist some semidefinite positive matrices $R_{i}(r)$ and $S_{i}(s)$ for separable Werner and isotropic states. It is obvious by noticing that $\vec{r}_{i}$s are Bloch vectors corresponding to SIC-POVM, that is, $\vec{r}_{i}$s are Bloch vectors of pure states. In light of \cref{Lemma-opposite}, $r,s\in \left[-\sqrt{\frac{2}{N(N-1)}} ,\sqrt{\frac{2(N-1)}{N}}\right]$ correspond to positive semidefinite density matrices.
Q.E.D.

The result of Theorem 1 is exhibited in \cref{Figure-contour}, where every point on the contour $rs=\tau$ provides a decomposition form for the state with parameter $\tau$.
Analogously, according to the parameter space for isotropic states in \cref{Figure-WI-parameters} and the relation between \cref{Werner-tau,ISO-tau}, the separable decomposition for isotropic states would be
\begin{align}
\rho_{\mathrm{I}}  = \sum_{i=1}^{N^2} \frac{1}{N^2} \left(\frac{1}{N}\mathds{1} + \frac{r}{2 |\vec{r}_i|} \vec{r}_i \cdot \vec{\lambda} \right) \otimes \left(\frac{1}{N}\mathds{1} + \frac{s}{2|\vec{r}_i|} \vec{r}_i \cdot \vec{\lambda}^{\,\mathrm{T}} \right) \; .
\end{align}
Consequently, in this way all the separable Werner and isotropic states are decomposed into product of local density matrices. To write out the decompositions we need to know the $N^2$ vectors $|\psi_i\rangle$ in SIC-POVM, and explicit constructions of the SIC-POVM in lower dimensional spaces can be found in \cite{Exp-SIC-POVM1,Exp-SIC-POVM2}.

As mentioned in above, the decomposition is generally not unique, i.e., for a given $\tau$ every point on the contour $rs=\tau$ in \cref{Figure-contour} corresponds to a separable decomposition. Every Bloch vector of the decomposed local density matrix matches to a SIC-POVM in $N$-dimensional Hilbert space. Note that though the existence of SIC-POVM in arbitrary dimension is still an open question, we find the following dimensions may yield definite solutions
\begin{eqnarray}
 N & = & 2,3,\cdots, 24, 28,30,31,35,37,39,43,48,124, \nonumber \\
  & & 143, 147, 168, 172, 195, 199, 228, 259, 323.
\end{eqnarray}
Readers may refer to Ref. \cite{SIC-Dim, SIC-review} for a recent review on this point.

\begin{figure}\centering
\scalebox{0.375}{\includegraphics{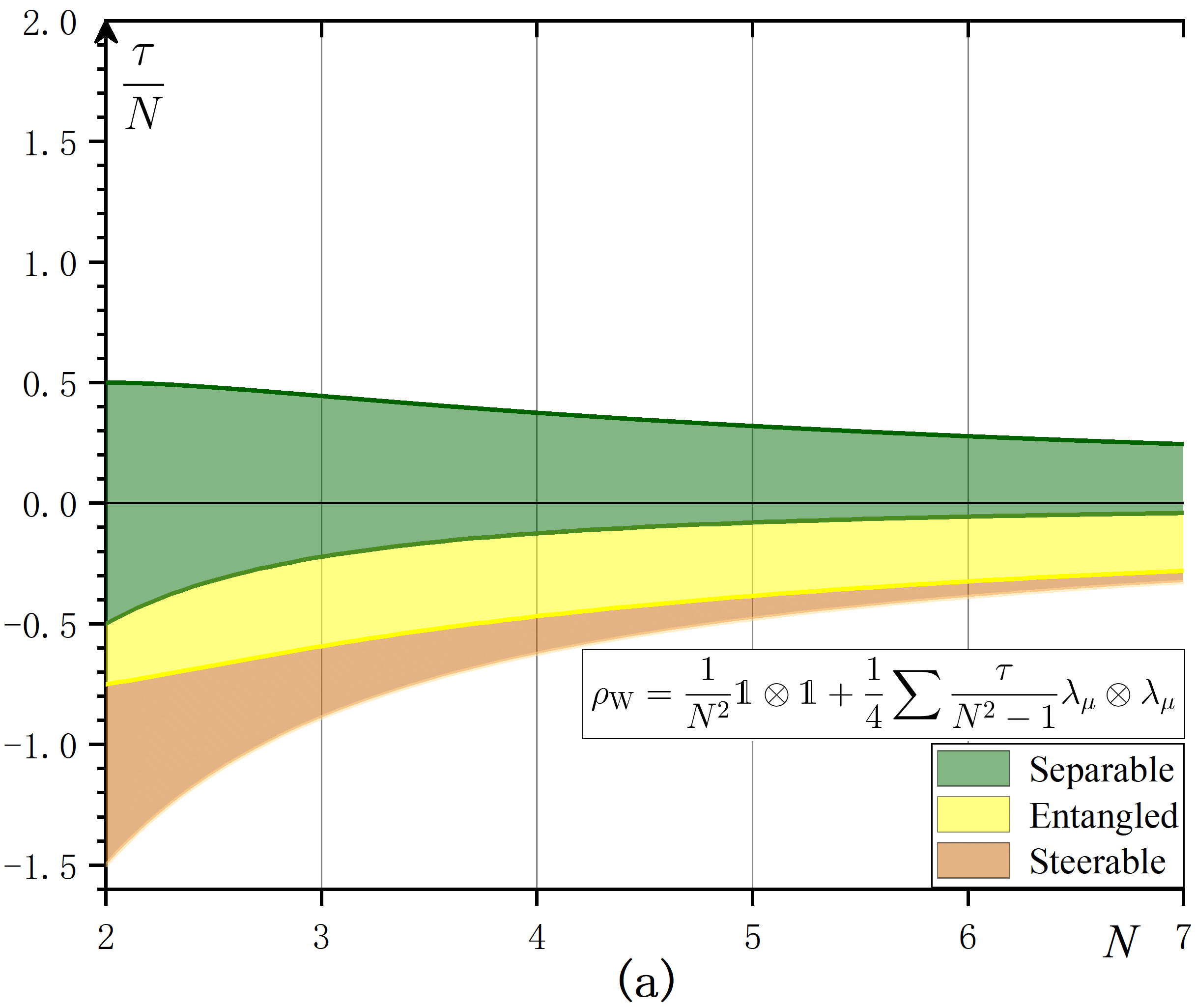}}\;
\scalebox{0.375}{\includegraphics{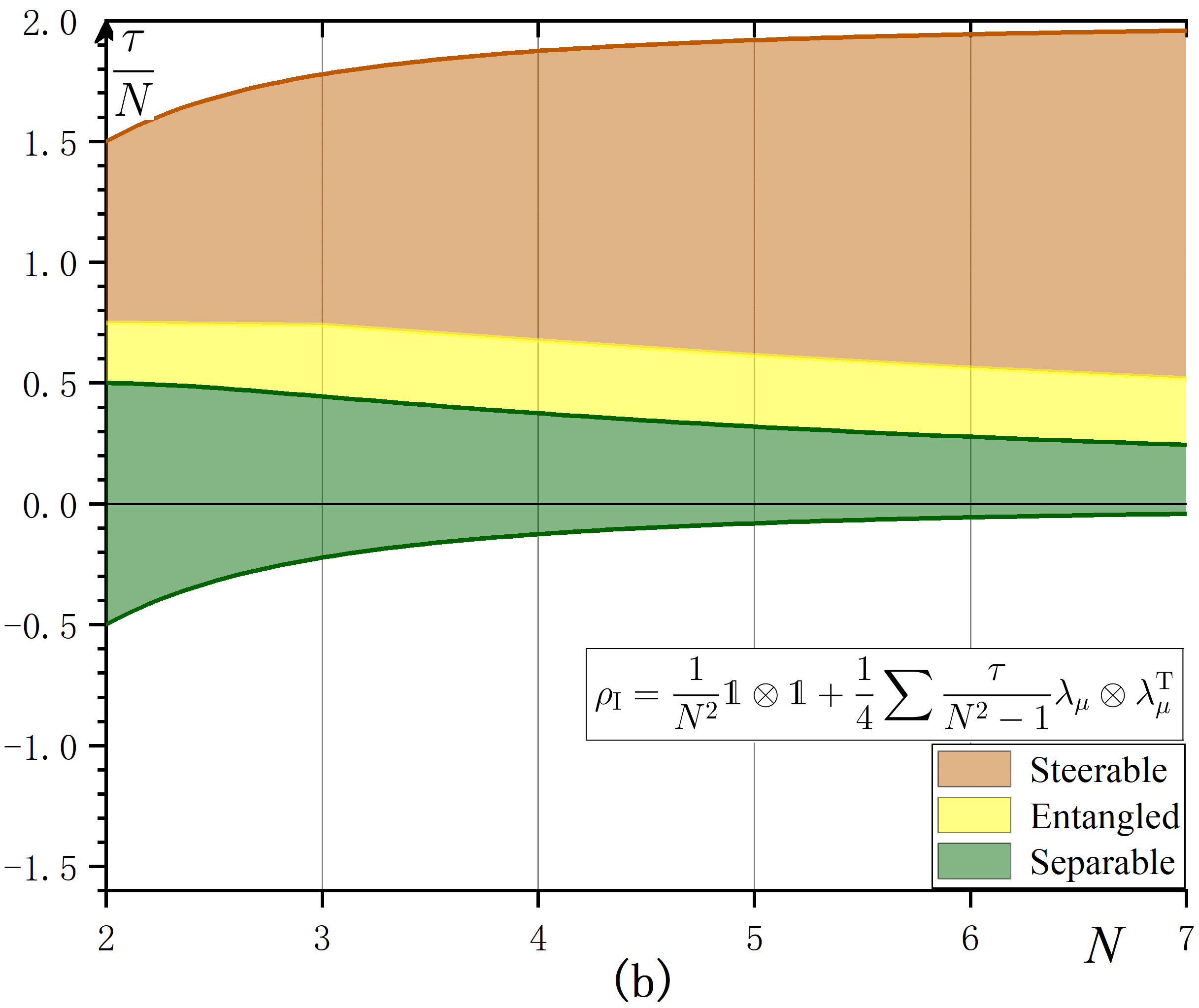}}
\caption{{\bf The nonlocalities of Werner and isotropic states in large $N$ limit.} Asymptotically, when dimension $N$ approaches to infinite, the Werner states $\rho_{\mathrm{W}}$ tend to be either separable or non-steerably entangled, while nearly all isotropic states $\rho_{\mathrm{I}}$ are steerable.}  \label{Figure-Asymp}
\end{figure}

The parameterization of Werner and isotropic states with single parameter $\tau$ enables us to study the asymptotic behaviour of large dimension $N$, see \cref{Figure-Asymp}. When dimension $N$ goes to infinity, in the parameter region of the Werner states there would be one half of separable states (the upper Green region in \cref{Figure-Asymp}(a)), and one half of non-steerable entangled states (the lower yellow region in in \cref{Figure-Asymp}(a)). While for isotropic states, the separable and non-steerable entangled states (green and yellow regions in \cref{Figure-Asymp}(b)) both become negligible in comparison with the steerable states (dark yellow regions in \cref{Figure-Asymp}(b)). The relative amount of the different types of non-locality may be well understood through the parametrization shown in \cref{Figure-WI-parameters}.

\section{Summary}

In this paper we proposed a novel scheme for the decomposition of all separable Werner and isotropic states in any dimension. To this aim, we first associated the decomposition of the separable Werner states with the existence of SIC-POVM. And then by taking advantage of the largest regular simplex in the Bloch-vector space, the separable Werner states and isotropic states are decomposed. With the unified parametrization scheme in Bloch vector representation, the Werner state were shown to be either separable or non-steerably entangled in large dimensions, asymptotically, while the isotropic states tend considerably to steerable states. That is to say that regarding quantum nonlocality the isotropic states exhibit more in both number and intensity than the Werner states.

\section*{Acknowledgements}
\noindent
This work was supported in part by the National Natural Science Foundation of China (NSFC) under the Grants 11975236 and 11635009; and by the University of Chinese Academy of Sciences.

\end{document}